\RequirePackage{fix-cm}
\documentclass[twocolumn]{svjour3}          
\smartqed  
\usepackage{graphicx}
\usepackage{mathptmx}      
\begin{document}

\title{Theoretical Modeling of the Non-equilibrium Amorphous State in 1T-TaS$_2$
}


\author{Jaka Vodeb$^{1,3}$           \and
        Viktor V. Kabanov$^{1}$      \and
        Yaroslav Gerasimenko$^{2}$   \and
        Igor Vaskivskyi$^{2}$        \and
        Jan Ravnik$^{1,3}$           \and
        Dragan Mihailovic$^{1,2,3}$ 
}


\institute{J. Vodeb \at
              \email{jaka.vodeb@ijs.si}           
           \and
              $^1$Jozef Stefan Institute, Jamova 39, 1000 Ljubljana, Slovenia \\
              $^2$CENN Nanocenter, Jamova 39, 1000 Ljubljana, Slovenia \\
              $^3$Dept. of Physics, Faculty for Mathematics and Physics, Jadranska 19, University of Ljubljana, 1000 Ljubljana, Slovenia
}

\date{Received: date / Accepted: date}

\maketitle

\begin{abstract}
1T-TaS$_2$ is known for it's remarkably complex phase diagram and it's unique long-lived metastable hidden (H) state. Recently, a novel metastable state has been discovered using higher fluences for photoexcitation than in the case of the H state. The state has been dubbed as amorphous (A) due to it's similarity to glass. Expanding on the work of Brazovskii and Karpov, we show that the A state can be successfully modeled with classical interacting polarons on a two dimensional hexagonal lattice. We have found that the polaron configuration of the A state corresponds to a frustrated screened Coulomb system, where there is no order-disorder phase transition.
\keywords{Charge density waves \and Polarons \and Lattice gas model \and Monte Carlo simulations}
\end{abstract}

\section{Introduction}
\label{intro}
The crystal of 1T-TaS$_2$ (TAS) exhibits a layered structure of S-Ta-S layers, where each of the atomic layers is arranged in a hexagonal lattice. Each Ta atom is surrounded by an octahedral arrangement of S atoms \cite{spijkerman1997}. Due to this highly anisotropic character of the crystal we may neglect any kind of three dimensional behavior and only focus on a single layer of TAS.

\begin{figure}
\centering
\includegraphics[scale=.84]{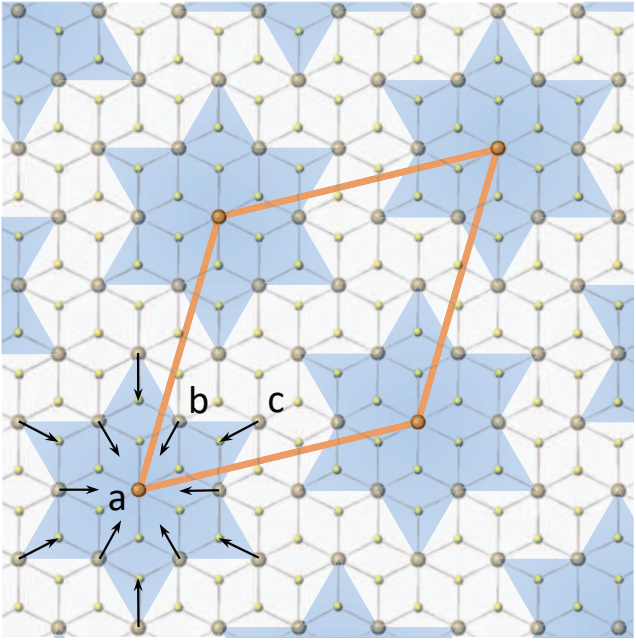}
\caption{\textbf{A schematic of the microscopic picture in the C phase:} A new unit cell is formed, where 12 Ta atoms are pulled towards the 13th in the center (marked with the letter a) in the shape of a star of David. The two different types of atoms with respect to their deviation from a perfect atomic lattice are marked with the letters b and c. The electron wave function also becomes localized on top of the central Ta atom. We interpret this unit cell as the polaron quasiparticle. The orange colored rhombus shows the unit cell of the hexagonal polaron lattice}
\label{ccdw}       
\end{figure}

The interplay of Coulomb interaction \cite{fazekas1979electrical}, spin-orbit coupling \cite{rossnagel2006spin} and electron-phonon interaction \cite{withers1986examination} results in various competing phases of TAS, which emerge as external parameters such as temperature and pressure are varied. When subjected to high pressure TAS becomes superconducting \cite{sipos2008}. At ambient pressure and high enough temperatures TAS is a homogeneous metal and upon lowering the temperature a series of charge density wave (CDW) transitions take place \cite{thompson1971transitions} as follows. At $540$ K an incommensurate CDW is formed, where the CDW modulation is incommensurate with underlying atomic lattice. When the temperature is lowered even further, at $350$ K, a nearly-commensurate CDW is formed, where a hexagonal structure of domains with a commensurate CDW is separated by domain walls, where the CDW is incommensurate with the atomic lattice. The phase below $190\sim220$ K has been interpreted as the commensurate CDW phase due to the perfect commensurability of the CDW with the underlying atomic lattice all throughout the crystal. The C phase also exhibits spin liquid like behavior \cite{klanjvsek2017}. Fig. \ref{ccdw} shows a schematic of the microscopic picture of the C phase.
In this paper we adopt the interpretation of the C phase as a system of classical interacting polarons first proposed by Brazovskii \cite{brazovskii2015}. Within this interpretation the ratio of the number of polarons and the number of atoms in the system or the \textit{polaron concentration} is exactly $1/13$. If polarons repel each other then they will arrange themselves in a hexagonal polaronic lattice, because this is the minimal energy configuration of such a system. The work of Stojchevska et al. \cite{stojchevska2014} has shown that the C phase can be photoexcited via an ultra-short laser pulse into a long-lived metastable state. Due to the fact that the excited phase cannot be found on the equilibrium phase diagram of TAS it has been dubbed as the hidden (H) state. Karpov et al. \cite{karpov2018} have subsequently shown by performing Monte Carlo simulations of a system of classical repulsive polarons that if the polaron concentration is perturbed slightly from $1/13$, the H state can be obtained within the scope of this interpretation. The classical Monte Carlo calculations neglect the quantum aspects of polarons and instead focus on the large scale configurational ordering of polarons. Quantum mechanical methods cannot do so due to their computational intensity. Methods used for analyzing the quantum mechanical aspects of polarons in the ground state include exact diagonalization \cite{wellein1997polaron}, variational exact diagonalization \cite{bonvca1999holstein}, quantum Monte Carlo \cite{de1983numerical}, variational methods \cite{davydov1979solitons} and DMRG \cite{jeckelmann1998density}. Other methods deal with the polaron dynamics and include Ehrenfest dynamics \cite{li2005ab,donati2016watching}, time-dependent density functional theory (TDDFT) \cite{polkehn2018quantum}, the multiconfiguration time-dependent Hartree method \cite{beck2000multiconfiguration}, its multilayer formulation \cite{wang2003multilayer} and the hierachical equations of motion \cite{tanimura1989time,chen2015dynamics}. TDDFT has been proven particularly successful in studying phenomena involved in polaron formation and dynamics, such as transient vibration energy redistribution and spectroscopy \cite{petrone2016ab} and defects in nanomaterials \cite{petrone2016quantum,stein2017cation}.

A recent paper by Gerasimenko et al. \cite{gerasimenko2018} has reported the discovery of a novel long-lived metastable state, where polarons are arranged in a disordered pattern. It is formed via photoexcitation with an ultrafast laser pulse, which confirms the state's non-equilibrium nature. It exhibits remarkable stability (at least 10 hours), which excludes any transient nature typically present in excited states. Moreover, even though the polaronic pattern suggests localization of charge, the system exhibits metallic behavior. Therefore the nature of this state is highly non-trivial and physically interesting. The main purpose of this paper is to show that the configurational aspect of the localized charges of this amorphous (A) state can also be understood within the scope of the classical polaron picture and to try to elucidate the reason for the disordered pattern.

\section{The Model}
\label{themodel}

The final A state can be modeled in the classical polaron picture of correlated electrons on a hexagonal lattice. TAS is known for the strong presence of the electron-phonon interaction \cite{rossnagel2011origin}. Therefore, we assume that equation $4.15$ in chapter $4.2$ in \cite{alexandrov1995polarons}, where small interacting polarons form in a system of electrons and phonons, is applicable. We neglect polaron hopping and spin, as we are only interested in the configurational aspect of the repulsive interaction between polarons. The Hamiltonian has the form
\begin{equation}
\mathcal{H}=\sum_{i,j}V(i,j)(n_i-\nu)(n_j-\nu),
\end{equation}
where $V(i,j)$ is the potential energy of interacting polarons, $n_i$ is $1$ if the site occupied by a polaron and $0$ otherwise,  and $\nu$ is the average polaron concentration. $\nu$ is defined as a ratio of the number of polarons in the system, which is fixed throughout the simulation, and the number of lattice sites in the finite lattice used in simulations. The interaction between polarons has two contributions. The attraction due to the overlap of phonon clouds and long range Coulomb repulsion.  The first part may be short or long-range depending on the type of phonons involved in polaron formation. If we assume that only polar phonons are responsible for polaron formation, the resulting interaction will be reduced to 
\begin{equation}
V(i,j)=\frac{e^2}{\epsilon_0\vert\textbf{r}_i-\textbf{r}_j\vert}
\end{equation}
where $\epsilon_0$ is the static dielectric constant of the material. Both the lattice and electrons contribute to it, however the lattice contribution is indeed dominant. The interaction with acoustic phonons as well as with nonpolar optical phonons leads to the short range attraction of polarons, but here we neglect the short range contribution to the potential. Since the system has finite conductivity the long range Coulomb repulsion is screened. Therefore, we approximate it by the Yukawa potential
\begin{equation}
V(i,j)=\frac{e^2}{\epsilon_0\vert\textbf{r}_i-\textbf{r}_j\vert}exp(-\vert\textbf{r}_i-\textbf{r}_j\vert/r_s),
\end{equation}
where $r_s$ is the screening radius.

\section{Monte Carlo Method}
\label{montecarlomethod}

Monte Carlo (MC) simulations of simulated annealing were performed using a standard Metropolis algorithm. A proposed successive state in the Markov chain consists of one polaron moving in a random direction and a finite distance away from its current position. Simulations were performed on a hexagonal lattice with the lattice constant $a$ and of the size of $84\times78$ sites, where periodic boundary conditions were imposed. The number of polarons in the system is constant throughout the simulation procedure. In order to simulate annealing, the system would always start off at a temperature of $10^{-2}$ measured in units of $e^2/\epsilon_0ak_B$ , which corresponds to the high temperature phase of the model for the commensurate (C) state. Afterwards, the system is cooled slowly down to a temperature of $10^{-3}$ with a temperature step of $10^{-5}$. There were $14\times10^6$ MC steps performed at each temperature step. One of the main reasons for introducing the Yukawa form of the polaron interaction is in the substantial growth of the MC relaxation time with the growth of the range of interaction. The screening length $r_s$ has been chosen as $4.5a$ and the interaction has been manually set to $0$ at distances greater than $24a$. At this value of $r_s$ the interaction range exceeds the radius of one polaron in C state ($\sqrt{13}a\approx3.61 a$). Furthermore, it is numerically feasible due to the manageable relaxation times, which follow from it. The code is available for reproducibility at \cite{MCcode}. 

\section{Tiling}
\label{tiling}

Tiling of a polaron configuration can serve as a good visual illustration of the effective charge density topography present in the system as well as all the different nearest neighbor distances present between polarons. The way both tilings are achieved is with the help of the so called Voronoi diagram. When a set of coordinates on a plane is given, which are the polaron coordinates, the plane can be partitioned into regions. Each coordinate is assigned a region of points in the plane, which are closer to it than to any other coordinate. This is done for the experimental amorphous state and the theoretical model for it in Fig. \ref{comparison}. Every Voronoi cell has one polaron charge associated with it and if we divide the charge by the cell’s area, we obtain the charge density of the cell. Since the charge of a polaron is fixed (at least in simulations), one can get a clearer perspective on the charge density present in the system just based on the coloring of all the different Voronoi cells. After the Voronoi diagram is complete, one can construct a different set of tiling polygons from the Voronoi diagram’s polygons. Each new polygon has 4 vertices and represents a distance between two neighboring coordinates. The first two vertices are the two neighboring coordinates from which the distance that the polygon represents is calculated. The other two vertices are the two points, which are part of the Voronoi polygons containing the two coordinates and are also the closest to both coordinates. This can be done for any two coordinates, which share two vertices of their Voronoi polygons. Because the distances between polarons in the model are discrete due to the perfect underlying hexagonal lattice, one can specify a color for each tiling polygon based on the distance between the two coordinates it connects. The coloring scheme for the experimental polaron coordinates is extended in the sense that the color spectrum in between any two colors represents the spectrum of all the possible distances between the two discrete distances.

\section{Results}
\label{results}

\begin{figure}
\centering
  \includegraphics[scale=.621]{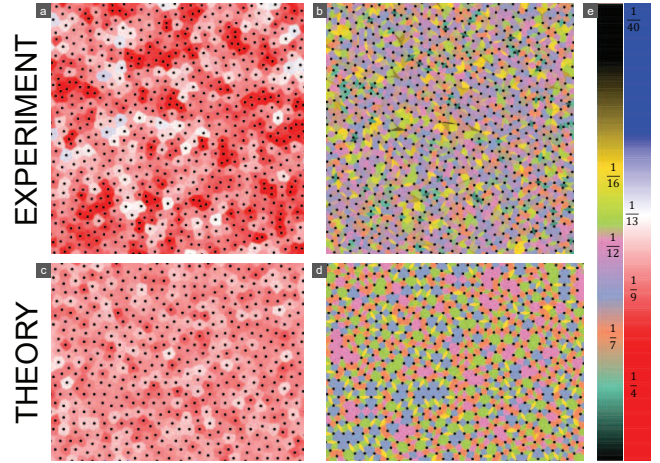}
\caption{\textbf{Comparison between the two types of tiling of experimental and theoretical A states:} \textbf{a}, Voronoi diagram of the experimental STM image of A state. The color in each cell represents the area of the cell according to the right legend in e. \textbf{b}, Nearest neighbor distance tiling of the experimental STM image of A state. The color of each tile represents the distance between the two polarons it connects according to the left legend in e. \textbf{c}, Voronoi diagram of the model for A state. \textbf{d}, Nearest neighbor distance tiling of the model for A state. \textbf{e}, Color legend for the tiling images to its left. Each color represents a different area of a Voronoi cell (right legend) and a different distance between nearest neighbors (left legend). Colors labeled with different polaron concentrations for the right legend represent areas of Voronoi cells of a Voronoi diagram, which is constructed from an overlying polaronic hexagonal lattice structure, which emerges at the corresponding concentration. For the legend to the left, colors represent different distances between nearest neighbors in the overlying polaronic hexagonal lattice structure at the corresponding polaron concentration}
\label{comparison}       
\end{figure}

\begin{figure}
\centering
  \includegraphics[scale=.5]{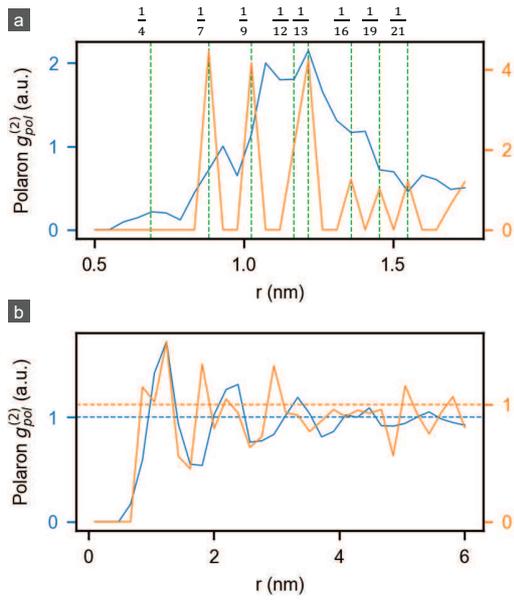}
\caption{\textbf{Comparison between the experimental (blue) and theoretical (orange) polaron pair correlation function $g_{pol}^{(2)}(r)$:} \textbf{a}, Comparison between experiment and theory, plotted only at small distances, where both are referenced to the atomic lattice constant. The green lines represent nearest neighbor distances of different polaron lattices, which emerge at different polaron concentrations of the system (from left to right: 1/4, 1/7, 1/9, 1/12, 1/13, 1/16, 1/19, 1/21).  The similarity between the theoretical model and experimental data can be observed from the qualitative matching of the polaron pair correlation peaks. The experimental peaks are almost all slightly shifted towards larger distances. However, the peaks of the theoretical model appear exactly at the expected positons and match the peaks of the experimental data. The quantitative comparison fails mostly at the 1/7 polaron concentration. \textbf{b}, Comparison between experiment and theory, plotted also at larger distances, where both are referenced to the atomic lattice constant. Both polaron pair correlations tend to the value 1 at large distances, which indicates a lack of ordering at large distances. The difference between them is in the oscillations around the value 1, where the theoretical model oscillates a lot more, which is due to the perfect underlying hexagonal lattice on which the polarons reside}
\label{g2pol}       
\end{figure}

\begin{figure}
\centering
  \includegraphics[scale=.45]{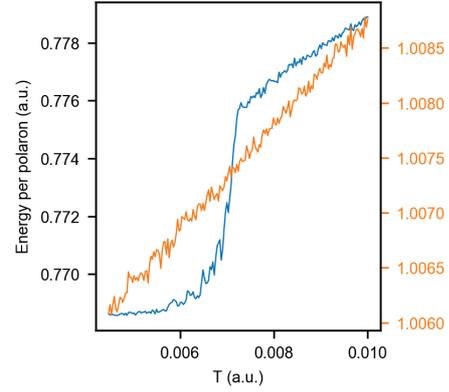}
\caption{\textbf{Energy of the system per polaron versus temperature for the model for C state (Blue) and A state (Orange):} The plot clearly shows the difference between the two model systems of polarons. A first order phase transition from disorder to order occurs in the case of the model for C state and no such phenomenon is present in the case of the model for A state. Energy is measured in units of $e^2/\epsilon_0ak_B$ and the error bars are smaller than the size of a single point}
\label{anneal_1}       
\end{figure}

\begin{figure}
\centering
  \includegraphics[scale=.57]{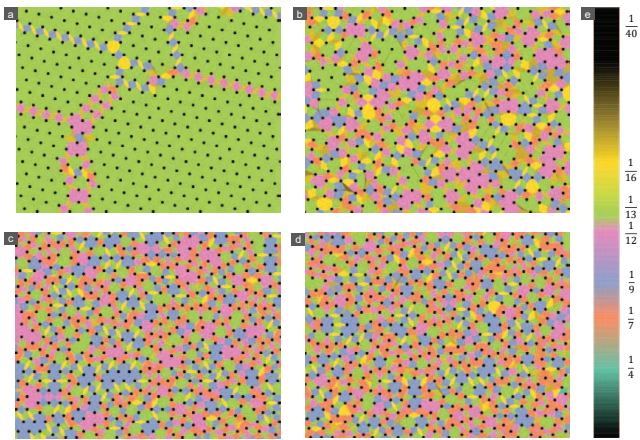}
\caption{\textbf{An examination of the difference between the theoretical models for C and A states:} \textbf{a}, \textbf{b}, Tiling of typical low and high temperature polaron configurations, respectively, in the model for C state. Clearly, the high temperature phase is highly disordered, while only one nearest neighbor distance prevails globally at low temperatures. There are a few line defects present still, but this is only due to finite size effects. \textbf{c}, \textbf{d}, Tiling of typical low and high temperature polaron configurations, respectively, in the model for A state. The high temperature phase is again clearly disordered. The low temperature phase, unlike in the case of the model for C state is still highly disordered. There are some nucleation centers of different nearest neighbor distances, which are not present at high temperatures, but it is obvious that the emergence of a global polaron lattice is prohibited. This is due to the commensurability frustration mentioned in the main text. \textbf{e}, Color legend for the nearest neighbor distances tiling images to its left. Each color represents a different distance between nearest neighbors, which corresponds to a listed polaron concentration in the legend}
\label{anneal_2}       
\end{figure}

Within the scope of the particular lattice gas model used in our simulations, the C phase can be understood as the ground state of a system of polarons, where the polaron concentration is exactly $1/13$. Experimental measurements of the polaron density in the STM images of the C and amorphous (A) states indicate that the polaron density in A state is about $1.23$ times higher than in C state \cite{gerasimenko2018}. Simulations were therefore performed at the corresponding polaron concentration and the result shows similarity with experimental data. Fig. \ref{comparison} shows the comparison between experiment and theory.
The takeaway from Fig. \ref{comparison} is the fact that both the theoretical and experimental polaron pattern exhibit disorder. A clearer comparison may be achieved via the polaron pair correlation function $g_{pol}^{(2)}(r)=\langle n(r)n(0)\rangle$.
Fig. \ref{g2pol} shows the comparison between the experimental data and theoretical modeling of A state within the scope of $g_{pol}^{(2)}(r)$.
The behavior of the theoretical model’s polaron pair correlation function is qualitatively very similar to the experimental behavior. Consider first Fig. \ref{g2pol}a. The green dashed lines mark all the possible lattice constants of polaronic lattices which are commensurate with the hexagonal atomic lattice. From the orange curve it is obvious that the simulation exhibits peaks in these cases. In experiment however, there is a discrepancy in the position of the peaks from the theoretically expected green dashed lines. This can be explained if the underlying atomic lattice has moved slightly in order to accommodate the repulsion between polarons. The qualitative discrepancy in peak intensity is mostly present at the $1/7$ polaron concentration peak, which is much more pronounced than in the experiment. Both of these discrepancies may vanish by incorporating also the movement of the atomic lattice in the model. This is the goal of future studies. In Fig. \ref{g2pol}b we can also see a discrepancy between the experimental and simulated peak positions. This is expected already from Fig. \ref{g2pol}a, because of the homogeneity of the sample. The behavior at close distances is reflected in long range behavior. Ergo, if there was a discrepancy in the peak positions at close range, it will also be present at long range. The main point of Fig. \ref{g2pol}b is to show that despite the peak position discrepancies, both the experimental as well as the theoretical curve tend to the value $1$ at large distances. This is a very important feature, as it shows that both exhibit no long range correlations. Within the scope of the theoretical model, the difference between C and A state can be understood by taking a look at a typical annealing process. Fig. \ref{anneal_1} shows the energy of the system per polaron dependence on temperature for models of both states measured in units of $e^2/\epsilon_0ak_B$.
Fig. \ref{anneal_2} shows typical configurations of both models at high and low temperatures.
It points out a clear difference in the two low temperature phases of the model for C state (Fig. \ref{anneal_2}a,b) and A state (Fig. \ref{anneal_2}c,d). The high temperature configuration in both cases (Fig. \ref{anneal_2}b,d) is disordered, yet the C state shows a clear strong tendency towards crystallization at low temperatures evident by the dominance of the green color (Fig. \ref{anneal_2}a). The A state (Fig. \ref{anneal_2}c) is clearly still disordered even at low temperatures. In the case of the model for C state, there is a clear order-disorder phase transition, which is of the first order. Meanwhile, there is no phase transition present in the case of the model for A state. In the model for A state in Fig. \ref{anneal_2}c some aggregation of the same types of different nearest neighbor distances can be observed at low simulated temperatures, but no global lattice ever emerges like in the model for C state. 

\section{Discussion}
\label{discussion}

We have clearly shown that the A state can be modeled successfully within the scope of the classical polaron picture. This result supports the model for the low temperature C state first proposed by Brazovskii \cite{brazovskii2015}. This model was also successfully used by Karpov et al. \cite{karpov2018} for simulating the H state. Our results imply that the A state's disordered characteristics are a consequence of commensurability frustration. The polaron hexagonal lattice at this polaron concentration cannot form due to the incommensurability with the underlying atomic lattice. One might argue that phase separation, where multiple commensurate lattices emerge in the system with domain walls in between them, might be energetically more favorable that the disordered phase seen in Fig. \ref{anneal_2}c. We have shown that this does not happen. Similar behavior has been observed before on a square lattice and it was also attributed to frustration with the underlying lattice \cite{lee2002,rademaker2013}. Minimization of the overlying hexagonal lattice energy on an underlying hexagonal lattice was done by Pokrovsky et al. \cite{pokrovsky1978}. They have shown that multiple minima very close in energy exist at integer polaron concentrations between $1/12$ and $1/9$ as well as between other concentrations. However, it is still not clear whether phase separation would result in further minimization of the energy of the system and whether unscreened interactions would change the system's behavior. Therefore, the observed glassy behavior may merely be a consequence of the limitations of the Monte Carlo algorithm, or perhaps the glassy behavior is an actual thermodynamic property of the system at this polaron concentration.

\section{Conclusion}
\label{conclusion}

In this paper the newly discovered meta-stable amorphous state in 1T-TaS$_2$ has been successfully modeled within the scope of a classical gas of polarons. The disordered structure has been found to be a consequence of frustration in the system. The hexagonal lattice of the polaron gas which seeks to emerge due to the repulsive interaction between them in not commensurate with the hexagonal underlying atomic lattice. According to our Monte Carlo simulations the result of this frustration is a remarkable stability of a disordered polaronic pattern, where there is no evidence of a first order crystallization phase transition. Our finding represent a new and important step in understanding photoexcitation in the complex system of 1T-TaS$_2$. The external laser pulse causes the system to form a greater number of polarons than in the thermodynamically stable commensurate state. Afterwards, polarons order in compliance with the screened interaction between them and the atomic lattice underneath into a stable disordered pattern.


%
%

\begin{acknowledgements}
We wish to thank Tomaz Mertelj for the useful discussions. The work was supported by ERC-2012-ADG\textunderscore 20120216 ``Trajectory'' and the Slovenian Research Agency (program P1-0040 and young researcher P0-8333).
\end{acknowledgements}

\bibliographystyle{spphys}       
\bibliography{Bibliografija}   

%
%

\end{document}